\documentclass[a4paper,reqno]{amsart}
\usepackage{graphicx}
\usepackage{amssymb}
 \footskip=30pt \numberwithin{equation}{section}
\begin{document}
\title[]{Inequivalent representations of commutator or anticommutator rings of
field operators and their applications}
\author{M. Matejka\\\\\emph{Department of Chemical Physics, Faculty of Mathematics,
Physics and Computer Sciences,
Comenius University, Mlynsk{\' a} dolina, 842 48 Bratislava,
Slovakia}
\\\\M. Noga\\\\\emph{Department of Theoretical Physics, Faculty of Mathematics,
Physics and Computer Sciences,
Comenius University, Mlynsk{\' a} dolina, 842 48 Bratislava, Slovakia}}%
\maketitle
\begin{abstract}\label{abstr}
Hamiltonian of a system in quantum
field theory can give rise to infinitely many partition functions
which correspond to infinitely many inequivalent representations
of the canonical commutator or anticommutator rings of field
operators. This implies that the system can theoretically exist
in infinitely many Gibbs states. The system resides in the
Gibbs state which corresponds to its minimal Helmholtz free energy
at a given range of the thermodynamic variables. Individual
inequivalent representations are associated with different
thermodynamic phases of the system. The BCS Hamiltonian of
superconductivity is chosen to be an explicit example for the
demonstration of the important role of inequivalent
representations in practical applications. Its analysis from the
inequivalent representations' point of view has led to
a recognition of a novel type of the superconducting phase
transition.
\end{abstract}
PACS numbers: 03.70.+k, 05.30.-d, 11.10.-z, 74.20.Fg, 74.25.Bt,
74.78.Bz

\section{Introduction}\label{sec1}
In quantum field theories based on operator formalism,  the
creation and annihilation field operators
$a^{\dagger}_{\vec{k},\sigma}$ and $a_{\vec{k},\sigma}$ are the fundamental 
objects creating and annihilating resp. particles 
in quantum states denoted by quantum numbers
$(\vec{k},\sigma)$, as for example, by the momentum $\vec{k}$ and
the spin projection $\sigma$. In any quantum field theory the
number of the operator
$(a_{\vec{k},\sigma},a^{\dagger}_{\vec{k},\sigma})$ pairs is infinite.
These operators satisfy the canonical commutation or
anticommutation relations
\begin{eqnarray}\label{1.1}
\big\{a_{\vec{k},\sigma},a^{\dagger}_{\vec{k^{'}},\sigma'}\big\}=
\delta_{\vec{k},\vec{k}^{'}}\delta_{\sigma,\sigma'}\;,\;
\big\{a_{\vec{k}^{'},\sigma'},a_{\vec{k},\sigma}\big\}=
\big\{a^\dagger_{\vec{k}^{'},\sigma'},a^\dagger_{\vec{k},\sigma}\big\}=0
\;.
\end{eqnarray}
The operators $a_{\vec{k},\sigma},a^{\dagger}_{\vec{k},\sigma}$
act on state vectors $\psi$ which span a Hilbert space
$\mathcal{H}$. In order to achieve a unique specification of the
commutator or anticommutator ring of the operators (\ref{1.1}),
in addition to (\ref{1.1}) one requires the existence of a vacuum
state $\phi_0$ in the Hilbert space $\mathcal{H}$ for which
\begin{eqnarray}\label{1.2}
a_{\vec{k},\sigma}\phi_0=0
\end{eqnarray}
for all $(\vec{k},\sigma)$. In this case, the Hilbert space
$\mathcal{H}$ is a space for a representation of the commutator
or anticommutator ring (\ref{1.1}) with the auxiliary condition
(\ref{1.2}).

As long as the number of the operators
$a_{\vec{k},\sigma},a^{\dagger}_{\vec{k},\sigma}$ entering the
algebraic structure (\ref{1.1}) and (\ref{1.2}) is finite, there
exists only one inequivalent representation for the algebraic
relations (\ref{1.1}) and (\ref{1.2}). However, in quantum field
theories describing systems with an infinite number of degrees of
freedom, the algebraic structure (\ref{1.1}) has infinitely many
inequivalent representations \cite{haag}. Intuitively speaking,
one can say that there exist infinitely many different and
inequivalent matrix realizations of the operators
$a_{\vec{k},\sigma}$ and $a^{\dagger}_{\vec{k},\sigma}$ satisfying
the same algebraic structure as (\ref{1.1}) and (\ref{1.2}). The
situation reminds us very distantly of a Lie algebra of a
non-compact group which has infinitely many unitary irreducible
representations for its generators realized in forms of infinitely
dimensional matrices. In contrast to the aforementioned Lie
algebra with a finite number of its generators, the canonical ring
(\ref{1.1}) involves infinite number of the elements
$a_{\vec{k},\sigma}$ and $a^{\dagger}_{\vec{k},\sigma}$ which can
be realized by infinitely many different and inequivalent
representations in forms of infinitely dimensional matrices.

The operators $a_{\vec{k},\sigma}$ and
$a^{\dagger}_{\vec{k},\sigma}$ entering the ring (\ref{1.1}) are
assumed to form a complete set of operators which means that every
operator in a quantum field theory can be built up out of them.
Thus a grand canonical Hamiltonian $H$ governing the dynamics of a
given physical system is expressed as a given function of
$a_{\vec{k},\sigma}$ and $a^{\dagger}_{\vec{k},\sigma}$, i.e.,
\begin{eqnarray}\label{1.3}
H=H(a^{\dagger},a)\;.
\end{eqnarray}
The grand canonical partition function $Z$ is expressed as the
density matrix trace
\begin{eqnarray}\label{1.4}
Z=Tre^{-\beta H}
\end{eqnarray}
where $\beta=1/T$ is the inverse temperature. All thermodynamic
properties of the system are determined by the grand canonical
potential
\begin{eqnarray}\label{1.5}
\Omega=-T\ln Z\;.
\end{eqnarray}
The statistical average values corresponding to physical
observables associated with operators $A(a^{\dagger},a)$ are
determined by the relations
\begin{eqnarray}\label{1.6}
\big<A\big>=\frac{1}{Z}Tr\{A(a^{\dagger},a)e^{-\beta H}\}\;.
\end{eqnarray}
The grand canonical potential (\ref{1.5}) and statistical average
values (\ref{1.6}) specify the so-called Gibbs state of the system
governed by the given Hamiltonian (\ref{1.3}).

Since the commutator or anticommutator ring (\ref{1.1}) admits
infinitely many inequivalent representations for the operators
$a_{\vec{k},\sigma}$ and $a^{\dagger}_{\vec{k},\sigma}$, it
implies that one has to associate the corresponding inequivalent
representations to the Hamiltonian $H$, partition function $Z$ and
grand canonical potential $\Omega$. Again, intuitively
speaking, one can say that the matrix form of the Hamiltonian
(\ref{1.3})  is distinct for each inequivalent representation of
the ring (\ref{1.1}). This implies that the partition function
(\ref{1.4}), grand canonical potential (\ref{1.5}) and statistical
average values of physical observables (\ref{1.6}) are distinct
for each inequivalent representation of (\ref{1.1}). In other
words, the same Hamiltonian $H$ gives rise to
different results for $Z$, $\Omega$ and $\big<A\big>$
corresponding to chosen inequivalent representations of the
canonical ring (\ref{1.1}). It implies that in each quantum field
theory a given Hamiltonian can give rise to infinitely many Gibbs
states. This theoretical conclusion seems to be in a conflict with
the experience because every physical system resides always in a single Gibbs
state for given values of thermodynamic variables, like temperature $T$, volume $V$
and particle number $N$ . The single Gibbs state corresponds to a single inequivalent
representation of the commutator or anticommutator ring
(\ref{1.1}).

The answer how one should select an appropriate single inequivalent
representation for a system at given values
$T$, $V$ and $N$ out of infinitively many representations is
uniquely given by the second law of thermodynamics. The second law
of thermodynamics requires, e.g., the Helmholtz free energy
$F(T,V,N)$ at given values of the thermodynamic variables $T$,
$V$ and $N$ to be minimal with respect to any free parameters
entering $F(T,V,N)$. Theoretically, it means that one should
evaluate the Helmholtz free energies $F(T,V,N)$ for all
inequivalent representations of (\ref{1.1}) and select that single
one which corresponds to their infimum at the given range of the
thermodynamic variables $T$, $V$ and $N$. A concrete example will demonstrate how 
it is done in practice.

For detailed understanding of inequivalent
representations of the commutator or anticommutator ring (\ref{1.1}) of field
operators' role in practical applications we will study them
from three different aspects. First, we will explicitly construct
a certain class of inequivalent representations of the
anticommutator ring (\ref{1.1}) of field operators. Second, we
will show how one, in practical calculations, tacitly selects a single 
inequivalent representation by choosing an
appropriate perturbation theory. At third, theoretical
implications of inequivalent representations of electron field operator' anticommutator
ring will be explicitly demonstrated
on the BCS model Hamiltonian of superconductivity \cite{bcs}. More specifically, 
a new class of inequivalent representations of (\ref{1.1}), which has not been
known till now and leads to a new unexpected superconducting state, will
be constructed.
\newpage

\section{Inequivalent representations}\label{sec2}
For the purpose of practical applications to the BCS model
Hamiltonian \cite{bcs} and for the sake of simplicity, we consider
a complete set of annihilation and creation operators
$a_{k,\sigma}$ and $a^{\dagger}_{k,\sigma}$ of the
fermion type. Let the index $k$ run over integer numbers
over the interval $k\in \big<-\frac{N}{2},\frac{N}{2}\big>$ and
$\sigma$ denotes spin $1/2$ projection of a fermion, i.e.,
$\sigma=\; \downarrow,\uparrow \;=+,-$. In order to have a quantum
field theory, we take the limit $N \to \infty$. The field
operators $a_{k,\sigma}$ and $a^{\dagger}_{k,\sigma}$
satisfy the anticommutator ring
\begin{eqnarray}\label{2.1}
\big\{a_{k,\sigma},a^{\dagger}_{k^{'},\sigma'}\big\}=
\delta_{k,k^{'}}\delta_{\sigma,\sigma'}\;,\;
\big\{a_{k^{'},\sigma'},a_{k,\sigma}\big\}=
\big\{a^\dagger_{k^{'},\sigma'},a^\dagger_{k,\sigma}\big\}=0
\;.
\end{eqnarray}
with the subsidiary condition
\begin{eqnarray}\label{2.2}
a_{k,\sigma}\phi_0=0
\end{eqnarray}
on the vacuum state $\phi_0$ for all $(k,\sigma)$. The
representation space for the anticommutator ring (\ref{2.1}) with
the subsidiary condition (\ref{2.2}) can be chosen to be the
Hilbert space $\mathcal{H}$ spanned by the basis vectors
$\psi_{\{n_{k,\sigma}\}}$ defined by the formula
\begin{eqnarray}\label{2.3}
\psi_{\{n_{k,\sigma}\}}=\lim_{N\to
\infty}\prod_{k,\sigma}\big(a^{\dagger}_{k,\sigma}\big)^{n_{k,\sigma}}
\phi_0\;,
\end{eqnarray}
where $n_{k,\sigma}=0,1$ are the occupation numbers of
fermions in states $(k,\sigma)$, and $\{n_{k,\sigma}\}$
denotes an infinite number array of items $0$ and $1$.
Each such infinite array specifies one of the basis vectors of the
Hilbert space $\mathcal{H}$.

Next we construct a class of inequivalent representations of the
anticommutator ring (\ref{2.1}) by adopting the same approach as
outlined in Haag's work \cite{haag}, however, for boson operators.
We start from the operators $a_{k,\sigma}$ and
$a^{\dagger}_{k,\sigma}$ obeying (\ref{2.1}) and introduce
the "unitary" transformations
\begin{eqnarray}\label{2.4}
\nonumber c_{k,\sigma} = e^{iQ}a_{k,\sigma}e^{-iQ}\;, \\
c^{\dagger}_{k,\sigma} = e^{iQ} a^{\dagger}_{k,\sigma}
e^{-iQ}\;,
\end{eqnarray}
where $Q$ is the Hermitian "operator"
\begin{eqnarray}\label{2.5}
Q=\lim_{N\to
\infty}\sum^{N/2}_{k=-N/2}\alpha_{k}T_{k}\;\;,\;\;
T_{k}=i\big(a^{\dagger}_{k,+}a^{\dagger}_{-k,-}-
a_{-k,-}a_{k,+}\big)
\end{eqnarray}
and $\alpha_{k}$ are arbitrary real parameters. The
anticommutation relations for the transformed operators
$c_{k,\sigma}$ and $c^{\dagger}_{k,\sigma}$ are, of
course, the same as given by (\ref{2.1}). The operator $e^{iQ}$
can be expressed as the following infinite product
\begin{eqnarray}\label{2.6}
e^{iQ}=\lim_{N\to \infty}\prod^{N/2}_{k=-N/2} \big[
1+iT_{k}\sin \alpha_{k}+T^2_{k}(\cos
\alpha_{k}-1) \big]
\end{eqnarray}
where
\begin{eqnarray}\label{2.7}
T^2_{k}=2a^{\dagger}_{k,+}a_{k,+}a^{\dagger}_{-k,-}
a_{-k,-}-a^{\dagger}_{k,+}a_{k,+}-a^{\dagger}_{-k,-}a_{-k,-}+1\;.
\end{eqnarray}
The transformations (\ref{2.4}), if evaluated through (\ref{2.4}),
are similar to the well-known Bogoliubov-Valatin transformations
\cite{bogval}
\begin{eqnarray}\label{2.8}
\nonumber
c_{k,+}=u_{k}a_{k,+}+v_{k}a^{\dagger}_{-k,-}\;\;,\;\;
c^{\dagger}_{k,+}=u_{k}a^{\dagger}_{k,+}+v_{k}a_{-k,-}
\\c_{k,-}=u_{k}a_{k,-}-v_{k}a^{\dagger}_{-k,+}\;\;,\;\;
c^{\dagger}_{k,-}=u_{k}a^{\dagger}_{k,-}-v_{k}a_{-k,+}\;,
\end{eqnarray}
where
\begin{eqnarray}\label{2.9}
\nonumber u_{k}=\cos \alpha_{k} \\
v_{k}=\sin \alpha_{k}\;.
\end{eqnarray}

In the limit $N \to \infty$, the operator $e^{iQ}$  given by
(\ref{2.4}) is not a proper operator but transforms every vector
$\psi$ of the Hilbert space $\mathcal{H}$ into one
$\psi^{'}=e^{iQ}\psi$ of a Hilbert space $\mathcal{H}^{'}$ with
unexpected properties explained as follows. Let us denote by
$\varphi_{\{n_{k}\}}$ any basis vector of $\mathcal{H}$
given by the formula
\begin{eqnarray}\label{2.10}
\varphi_{\{n_{k}\}}=\lim_{N \to
\infty}\prod^{N/2}_{k=-N/2}\big(a^{\dagger}_{k,+}
a^{\dagger}_{-k,-}\big)^{n_{k}}\phi_0
\end{eqnarray}
where $n_{k}=n_{k,+}=n_{-k,-}=0,1$. All the
basis vectors $\varphi_{\{n_{k}\}}$ form a subspace of
$\mathcal{H}$. The transformation $e^{iQ}$ transforms every basis
vector $\varphi_{\{n_{k}\}}$ into one
$\varphi'_{\{n^{'}_{k}\}}=e^{iQ}\varphi_{\{n_{k}\}}$
of $\mathcal{H}^{'}$, given by the formula
\begin{eqnarray}\label{2.11}
\varphi'_{\{n^{'}_{k}\}}=\lim_{N\to
\infty}\prod^{N/2}_{k=-N/2}\bigg\{\Big[\delta_{n_{k},1}-
\big(a^{\dagger}_{k,+}a^{\dagger}_{-k,-}\big)^{{n_{k}}+1}\Big]\sin
\alpha_{k}+\big(a^{\dagger}_{k,+}a^{\dagger}_{-k,-}\big)^{n_{k}}\cos
\alpha_{k}\bigg\}\phi_0\;.
\end{eqnarray}
The result is that the scalar product
$\Big(\psi,e^{iQ}\varphi_{\{n_{k}\}}\Big)$ for every basis
vector $\psi$ of $\mathcal{H}$ given by (\ref{2.3}) is either
identically equal to zero or equal to the infinite product
\begin{eqnarray}\label{2.12}
\Big(\psi_{\{n^{'}_{k^{'}}\}},e^{iQ}\varphi_{\{n_{k}\}}\Big)=\lim_{N\to
\infty}\prod^{N/2}_{k^{'},k=-N/2}\Big(S_{k,k^{'}}\sin
\alpha_{k}+C_{k,k^{'}}\cos
\alpha_{k}\Big)\;,
\end{eqnarray}
where
\begin{eqnarray}\label{2.13}
\nonumber S_{k,k^{'}}=\delta_{0,n^{'}_{k^{'}}}
\delta_{1,n_{k}}-\delta_{0,n_{k}}\;\delta_{1,n^{'}_{k^{'}}}
\delta_{k,k^{'}} \\
C_{k,k^{'}}=\delta_{n_{k},n^{'}_{k^{'}}}
\delta_{k,k^{'}}\;.
\end{eqnarray}
However, the infinite product (\ref{2.12}) diverges also to zero
in the limit $N \to \infty$. This is so because by noting the
properties of the coefficients $S_{k,k^{'}}$ and
$C_{k,k^{'}}$ given by (\ref{2.13}) we see that the
scalar product (\ref{2.12}) reduces to the infinite product of the
following type
\begin{eqnarray}\label{2.14}
\Big(\psi_{\{n^{'}_{k^{'}}\}},e^{iQ}\varphi_{\{n_{k}\}}\Big)=\lim_{N\to
\infty}\prod^{N/2}_{k^{'},k=-N/2}\sin \alpha_{k^{'}}\;\cos
\alpha_{k}=0\;,
\end{eqnarray}
The last relation implies that the Hilbert space $\mathcal{H}^{'}$
spanned by the transformed state vectors $\psi^{'}=e^{iQ}\psi$
contains a subspace of the state vectors $\varphi'_{\{n_{k}\}}$
which are orthogonal to every vector $\psi$ of $\mathcal{H}$. We
make the same conclusion as in Haag's work \cite{haag}:
$c_{k,\sigma}$ and $c^{\dagger}_{k,\sigma}$ given by
(\ref{2.4}) are operators satisfying the same canonical ring as
(\ref{2.1}), i.e.,
\begin{eqnarray}\label{2.15}
\big\{c_{k,\sigma},c^{\dagger}_{k^{'},\sigma'}\big\}=
\delta_{k,k^{'}}\delta_{\sigma,\sigma'}\;,\;
\big\{c_{k^{'},\sigma'},c_{k,\sigma}\big\}=
\big\{c^\dagger_{k^{'},\sigma'},c^\dagger_{k,\sigma}\big\}=0
\end{eqnarray}
but there is no proper unitary transformation connecting these two
operator systems. The operators
$\big(a_{k,\sigma},a^{\dagger}_{k,\sigma}\big)$ and
the operators
$\big(c_{k,\sigma},c^{\dagger}_{k,\sigma}\big)$ act in
two different Hilbert spaces $\mathcal{H}$ and $\mathcal{H}^{'}$
respectively. In Haag's terminology, they belong to inequivalent
representations of the same anticommutator ring (\ref{2.1}) or
(\ref{2.15}) of the field operators. It is a simple matter to show
that among all transformed vectors $\psi^{'}=e^{iQ}\psi$ there is
no one such as $\phi^{'}_0$ for which
\begin{eqnarray}\label{2.16}
c_{k,\sigma}\phi^{'}_0=0
\end{eqnarray}
for all $(k,\sigma)$. Next we supplement the new Hilbert
space $\mathcal{H}^{'}$ spanned by all state vectors
$\psi^{'}=e^{iQ}\psi$ with the vacuum state $\phi^{'}_0$ 
satisfying the auxiliary condition (\ref{2.16}).

Each inequivalent representation of the anticommutator ring
(\ref{2.15}) with the subsidiary condition (\ref{2.16}) is
specified by a chosen infinite set of parameters
$\alpha_{k}$ entering the transformations
(\ref{2.4})-(\ref{2.9}). Thus, the number of the inequivalent
representations of (\ref{2.15}) is indeed infinite. In
contradistinction to inequivalent representations of Lie algebras,
where the representations are specified by a finite number of
Casimir operator eigenvalues, the inequivalent representations of
the anticommutator ring (\ref{2.15}) of the field operators
$c_{k,\sigma}$ and $c^{\dagger}_{k,\sigma}$ are
specified by infinite set of parameters.

Our analysis of inequivalent representations of the anticommutator
ring of the field operators (\ref{2.1}), as presented above, can
be generalized in a straightforward way to any set of quantum
numbers $(k,\sigma)$ and to many other classes of
inequivalent representations both for fermions and bosons.

Next we consider a Hamiltonian $H$ governing a physical system
with an infinite number of degrees of freedom as a given function
of the operators $a_{k,\sigma}$ and
$a^{\dagger}_{k,\sigma}$, i.e.,
\begin{eqnarray}\label{2.17}
H=H(a^{\dagger},a)\;.
\end{eqnarray}
The field operators $a_{k,\sigma}$ and
$a^{\dagger}_{k,\sigma}$ and the corresponding Hamiltonian
(\ref{2.17}) act in the representation space which is the Hilbert
space $\mathcal{H}$ spanned by the basis vectors given by
(\ref{2.3}). In this representation we have the corresponding
partition function
\begin{eqnarray}\label{2.18}
Z=Tre^{-\beta H}
\end{eqnarray}
and statistical average values $\big<A\big>$ of physical
observables associated with operators $A(a^{\dagger},a)$ given by
the relations
\begin{eqnarray}\label{2.19}
\big<A\big>=\frac{1}{Z}Tr\big\{A(a^{\dagger},a)e^{-\beta
H}\big\}\;.
\end{eqnarray}

For each inequivalent representation of the anticommutator ring
(\ref{2.15}) we can construct the transformed Hamiltonian
\begin{eqnarray}\label{2.20}
\widetilde{H}(c^{\dagger},c)=e^{iQ}H(a^{\dagger},a)e^{-iQ}
\end{eqnarray}
in its normal form by employing the canonical anticommutator
relations (\ref{2.15}). The corresponding partition function
\begin{eqnarray}\label{2.21}
\widetilde{Z}=Tre^{-\beta \widetilde{H}(c^{\dagger},c)}
\end{eqnarray}
can be different from that given by (\ref{2.18}) because the
operators $H(a^{\dagger},a)$ and $\widetilde{H}(c^{\dagger},c)$
act in two different Hilbert spaces $\mathcal{H}$ and
$\mathcal{H}^{'}$ respectively. By the same way, the statistical
average value $\big<A\big>$ corresponding to a physical observable
\begin{eqnarray}\label{2.22}
\big<A\big>=\frac{1}{\widetilde{Z}}Tr\big\{\widetilde{A}(c^{\dagger},c)e^{-\beta
\widetilde{H}}\big\}\;.
\end{eqnarray}
can be different from that given by (\ref{2.19}). These
conclusions may seem to be paradoxes residing in facts as if
physical observables were dependent on our will how we select a
single inequivalent representation for the anticommutator ring of
the field operators (\ref{2.15}). There is, in fact, no freedom in
selecting an appropriate inequivalent representation of
(\ref{2.15}). As we have mentioned in the introduction, the second
law of thermodynamics dictates uniquely which inequivalent
representation is relevant for describing the physical system at
given values of thermodynamic variables.

One may even incorrectly believe that the conclusion concerning
the different results for the same physical quantities given by
the relations (\ref{2.18}) and (\ref{2.21}) or by the relations
(\ref{2.19}) and (\ref{2.22}) are wrong. Such an incorrect belief
is supported by the formal appearance of the transformation
(\ref{2.20}) which seems to be a similarity transformation. The
unpermited application of the cyclic properties for the trace of 
infinitely dimensional matrices product would lead to incorrect
conclusions that the results of the relations (\ref{2.18}), 
(\ref{2.20}) or (\ref{2.19}), (\ref{2.21}) are
identical.

We conclude this section by stating that in quantum field theory 
a given Hamiltonian $H$ of a system leads to distinct results for
the partition function $Z$ and for the statistical average values
of physical observables $\big<A\big>$ depending on a selected
inequivalent representation of the anticommutator ring of field
operators. These conclusions will be explicitly demonstrated on
the BCS model Hamiltonian of superconductivity.
\newpage

\section{One method for the selection of individual inequivalent representations}\label{sec3}

In this section we elucidate how one tacitly selects a single
inequivalent representation out of infinitely many representations
of the commutator or anticommutator ring (\ref{1.1}) of field
operators in a practical application of quantum field theory. In
quantum field theories with interactions between fields there is
not known even one physical example with an exact solution. In all
practical applications one divides Hamiltonian $H$ of a system
into the sum
\begin{eqnarray}\label{3.1}
H=H_0(a^{\dagger},a)+H_I(a^{\dagger},a)
\end{eqnarray}
where $H_0$ is called the unperturbed Hamiltonian and the
remaining term $H_I$ is called the perturbative part. The
unperturbed Hamiltonian $H_0$ is chosen in a way to
be exactly diagonalized and by this fact its effects are treated
exactly. Its eigenvalues $\psi_{\mu}$, where $\mu$ denotes an
array with an infinite number of items form a complete basis of a
Hilbert space $\mathcal{H}$. The Hilbert space $\mathcal{H}$ is
the representation space for a single inequivalent representation
of the commutator or anticommutator ring (\ref{2.1}) of the field
operators $a_{k,\sigma}$ and $a^{\dagger}_{k,\sigma}$
entering the Hamiltonian (\ref{3.1}). The unperturbed partition
function
\begin{eqnarray}\label{3.2}
Z_0=Tre^{-\beta H_0(a^{\dagger},a)}
\end{eqnarray}
can be exactly evaluated and is typical for the chosen
inequivalent representation. The total partition function $Z$ is
expressed by the perturbation series
\begin{eqnarray}\label{3.3}
Z=Tre^{-\beta H}=Z_0\bigg<T\exp \bigg\{-\int^{\beta}_0d\tau
V(\tau)\bigg\}\bigg>_0=Z_0\sum^{\infty}_{\nu=1}\frac{(-1)^{\nu}}{\nu
!}\bigg<T\bigg(\int^{\beta}_0 d\tau V(\tau)\bigg)^{\nu}\bigg>_0
\end{eqnarray}
where the symbol $T$ stands for the time-ordered product,
\begin{eqnarray}\label{3.4}
V(\tau)=e^{\tau H_0}H_Ie^{-\tau H_0}
\end{eqnarray}
and
\begin{eqnarray}\label{3.5}
\bigg<T\bigg(\int^{\beta}_0 d\tau V(\tau)\bigg)^{\nu}\bigg>_0
\end{eqnarray}
denotes the statistical average value of the operator inside the
brackets $\big<...\big>_0$ with respect to the unperturbed
Hamiltonian $H_0$. It is needless to say that the statistical average
values (\ref{3.5}) are evaluated in the chosen inequivalent
representation.

However, the splitting of the total Hamiltonian as given by
(\ref{3.1}) is not unique. One can equally well divide the same
Hamiltonian as
\begin{eqnarray}\label{3.6}
H=H^{'}_0(a^{\dagger},a)+H^{'}_I(a^{\dagger},a)
\end{eqnarray}
where the new unperturbed Hamiltonian $H^{'}_0$ is by definition
different from $H_0$ and is not related to $H_0$ by any proper
unitary transformation. The new unperturbed Hamiltonian $H^{'}_0$
is assumed to be diagonalized by such transformations of the field 
operators $a_{k,\sigma}$ and $a^{\dagger}_{k,\sigma}$ like
(\ref{2.4})  in order to achieve its diagonal form
\begin{eqnarray}\label{3.7}
\widetilde{H}_0(c^{\dagger},c)=e^{iQ}H^{'}_0(a^{\dagger},a)e^{-iQ}\;.
\end{eqnarray}
The requirement to have $\widetilde{H}_0(c^{\dagger},c)$ in its
diagonal form puts certain constraints on the transformation
parameters $\alpha_{k}$ entering the transformations
(\ref{2.4})-(\ref{2.9}). In other words, all 
transformation parameter $\alpha_{k}$ values are determined by a
finite set of physical parameters present in the chosen
unperturbed Hamiltonian $H^{'}_0(a^{\dagger},a)$. Since the
infinite set of parameters $\alpha_{k}$ specifies a single
inequivalent representation of the commutator or anticommutator
ring of the field operators (\ref{2.15}), with new chosen
unperturbed Hamiltonian $H^{'}_0$ one again tacitly selects
another single inequivalent representation. The eigenvalues
$\psi'_{\mu}$ of $\widetilde{H}_0$ form again a complete basis of
a new Hilbert space $\mathcal{H}^{'}$ for the new selected
inequivalent representation.

In this inequivalent representation one gets the unperturbed
partition function
\begin{eqnarray}\label{3.8}
\widetilde{Z}_0=Tre^{-\beta \widetilde{H}_0(c^{\dagger},c)}
\end{eqnarray}
which is, of course, different from (\ref{3.2}) by the definitions
(\ref{3.1}) and (\ref{3.6}). The corresponding total partition
function
\begin{eqnarray}\label{3.9}
\widetilde{Z}=Tre^{-\beta
\widetilde{H}(c^{\dagger},c)}=\widetilde{Z}_0\sum^{\infty}_{\nu=0}\frac{(-1)^{\nu}}{\nu
!}\bigg<T\bigg(\int^{\beta}_0 d\tau
\widetilde{V}(\tau)\bigg)^{\nu}\bigg>_0
\end{eqnarray}
is also distinct from that given by (\ref{3.3}) because the same
Hamiltonian $H$ has two different, so to speak, "matrix"
realizations (\ref{3.1}) and (\ref{3.3}) corresponding to two
different inequivalent representations of the anticommutator ring
of field operators (\ref{2.1}) or (\ref{2.15}).

By the same approach as outlined above, one can continue to study
a series of inequivalent representations associated with a given
Hamiltonian $H$. By selecting a series
$H_0=H_{01},H_{02},H_{03},...$ of unperturbed Hamiltonians
$H_{01}$, $H_{02}$, $H_{03},...$ one can explore physical
properties of the corresponding series of different Gibbs states
associated with the same Hamiltonian $H$.

The aforementioned methods will be explicitly demonstrated for the
investigation of physical properties of several distinct Gibbs
states associated with the BCS model Hamiltonian in theory of
superconductivity \cite{bcs}. The BCS Hamiltonian has been
deliberately chosen from three different reasons.

At first, the BCS Hamiltonian is generally very well-known, extraordinary
simple and despite of its simplicity it has been very successful
in explaining properties of a large class of the so-called low
temperature superconductors (LTS) on terms of only two
phenomenological material parameters. On the other hand,
properties of high temperature superconductors (HTS) having
anisotropic layered structures are claimed to be unexplainable on
the basis of the BCS theory.

At second, the BCS theory as a theory with a nontrivial
interactions between electrons has provided us with two different
solutions for the Gibbs states which are asymptotically exact in
the thermodynamic limit.

At third, the BCS theory has been studied from various aspects 
for many years. That is why there is a generally spread belief
that its consequences are completely exhausted and therefore there
are no motivations for its further exploration. However, its
consequences has not been studied from the point of view of
inequivalent representations of the anticommutator ring of
electron field operators (\ref{2.1}) or (\ref{2.15}) yet. Since
the ring has infinitely many inequivalent representations the BCS
Hamiltonian should have infinitely many Gibbs states. It is,
therefore, worthwhile to undertake the task to find at least one
new Gibbs state associated with the BCS Hamiltonian.

\newpage

\section{The applications on the BCS theory of superconductivity}\label{sec4}
The BCS theory is based on the following well-known grand
canonical Hamiltonian
\begin{eqnarray}\label{4.1}
\nonumber H = \sum_{\vec{k},\sigma}
\xi_{\vec{k}}a^{\dag}_{\vec{k},\sigma}a_{\vec{k},\sigma}
-\frac{g}{V} \sum_{\vec{k}, \vec{k}^{'}}\,
a^{\dagger}_{\vec{k},+}a^{\dagger}_{-\vec{k},-}
a_{-\vec{k}^{'},-}a_{\vec{k}^{'},+}\Theta(\hbar\omega_{D} - \mid
\xi_{{\vec
k}}\mid ) \Theta(\hbar\omega_{D} - \mid \xi_{{\vec{k}^{'}}}\mid )\equiv \\
\equiv K+H_I\;,
\end{eqnarray}
where $\xi_{\vec{k}}=\frac{\hbar^2}{2m}\vec{k}^2-\mu$ is the
kinetic energy of an electron (in the state specified by the wave
vector $\vec{k}$) counted from the chemical potential $\mu$ which
can be approximated by the Fermi energy $\epsilon_F$, i.e., $\mu
\doteq \epsilon_F$. The index $\sigma = \pm$ denotes the spin
$1/2$ projection of an electron, the symbols $a_{\vec{k},\sigma}$
and $a^\dagger_{\vec{k},\sigma}$ are annihilation and creation
operators of electrons in the states $\big(\vec{k},\sigma \big)$
resp. Finally, $g$ is the squared electron-phonon coupling
constant, $\omega_D$ is the Debye frequency, $V$ is the volume of
the system. $K$ denotes the electron kinetic
energy operator and $H_I$ is the interaction term. The sum over $\vec{k}$
and $\vec{k}^{'}$ in $H_I$ is restricted by the conditions $\mid
\xi_{\vec{k}} \mid < \hbar \omega_D$, $\mid \xi_{\vec{k}^{'}} \mid
< \hbar \omega_D$, as indicated by the appropriate step functions
$\Theta(x)$ in the relation (\ref{4.1}).

All thermodynamic properties of the system governed by the
Hamiltonian (\ref{4.1}) are determined by the grand canonical
partition function
\begin{eqnarray}\label{4.2}
Z=Tre^{-\beta H}
\end{eqnarray}
The grand canonical potential
\begin{eqnarray}\label{4.3}
\Omega=-k_BT \ln Z
\end{eqnarray}
specifies the Gibbs state of the system and provides us with
complete information on thermodynamic properties of the system
described by the Hamiltonian (\ref{4.1}). $\Omega$ is a function
of the thermodynamic variables $T$, $V$ and $\mu$, and in addition
to them, it is also a function of the coupling constant $g$, i.e.,
$\Omega \equiv \Omega(T,V,\mu;g)$. From the definitions
(\ref{4.2}) and (\ref{4.3}) it follows
\begin{eqnarray}\label{4.4}
\frac{\partial \Omega}{\partial g}=\frac{1}{g}\big<H_I\big> \;,
\end{eqnarray}
where the statistical average $\big<H_I\big>$ is, of course, a non
trivial function of $g$, i.e., $\big<H_I\big>=\big<H_I\big>(g)$.
Suppose that one has calculated $\big<H_I\big>(g)$, then he can
integrate the equation (\ref{4.4}) to get the exact relation
\begin{eqnarray}\label{4.5}
\Omega-\Omega_n=\int_0^g \frac{dg'}{g'}\big<H_I\big>(g') \;,
\end{eqnarray}
where $\Omega \equiv \Omega(T,V,\mu;g)$ is the grand canonical
potential as defined by (\ref{4.2}) and (\ref{4.3}) and
$\Omega_n=\Omega(T,V,\mu;g=0)$ is the grand canonical potential
corresponding to the ideal electron gas. The grand canonical potential 
difference (\ref{4.5}) is particularly convenient
for two reasons. First, it expresses physical anomalies
associated with the interaction term $H_I$ above the background
corresponding to properties of the ideal electron gas. Second,
it exhibits manifestly the dependence of $\Omega$ on a chosen
inequivalent representation of the anticommutator ring (\ref{2.1})
of the electron field operators entering the Hamiltonian $H$. In
practical calculations one cannot evaluate the exact average value
$\big<H_I\big>$ and must resort to a perturbation theory by
selecting an unperturbed Hamiltonian $H_0$ which can be exactly
diagonalized.

With a chosen unperturbed Hamiltonian $H_0$ one calculates the
unperturbed grand canonical partition function (\ref{3.2}) and the
corresponding statistical averages $\big<A\big>_0$ of operators
$A$ defined by the relations
\begin{eqnarray}\label{4.6}
\big<A\big>_0=\frac{1}{Z_0}Tr\big(Ae^{-\beta H_0}\big)\;.
\end{eqnarray}
The average value $\big<H_I\big>(g)$ entering the relations
(\ref{4.4}) and (\ref{4.5}) can be expressed in the form
\begin{eqnarray}\label{4.7}
\big<H_I\big>(g)=\big<H_I\big>_0(g)+R(g) \;,
\end{eqnarray}
where $R(g)$ represents symbolically all contributions coming from
remaining perturbative terms.

First we choose $H_0$ to be the electron kinetic energy operator 
in (\ref{4.1}), i.e.,
\begin{eqnarray}\label{4.8}
H_{01}=\sum_{\vec{k},\sigma}
\xi_{\vec{k}}a^{\dag}_{\vec{k},\sigma}a_{\vec{k},\sigma} \;.
\end{eqnarray}
In this selection, the unperturbed Hamiltonian $H_{01}$ is already
diagonalized and the complete operator system is given by the
electron annihilation and creation operators $a_{\vec{k},\sigma}$
and $a^\dagger_{\vec{k},\sigma}$ resp. in the quantum states
$\big(\vec{k},\sigma\big)$. These operators obey the canonical
anticommutator ring (\ref{2.1}) with the subsidiary condition
(\ref{2.2}). In this case the Hilbert space $\mathcal{H}_1$ is the
space of a single inequivalent representation of the
anticommutator ring (\ref{2.1}) with the auxiliary condition
(\ref{2.2}). Each basis vector $\psi$ is then specified by an
infinite array $\big\{n_{\vec{k},\sigma}\big\}$ of the occupation
numbers $n_{\vec{k},\sigma}=0,1$ for each one particle state
$\big(\vec{k},\sigma \big)$, i.e.,
\begin{eqnarray}\label{4.9}
\psi\big(\big\{n_{\vec{k},\sigma}\big\}\big)=\prod_{\vec{k},\sigma}
\big(a^{\dagger}_{\vec{k},\sigma}\big)^{n_{\vec{k},\sigma}}\phi_0
\;.
\end{eqnarray}

With this choice of the unperturbed Hamiltonian $H_{01}$ one gets
the spectrum $E_{\vec{k},\sigma}$ of elementary excitations
\begin{eqnarray}\label{4.10}
E_{\vec{k},\sigma}=\xi_{\vec{k}} \;, \; \xi_{\vec{k}} \in
<-\mu,\infty)
\end{eqnarray}
and the statistical average values
\begin{eqnarray}\label{4.11}
\big<a_{\vec{k}^{'},\sigma'},a_{\vec{k},\sigma}\big>=
\big<a^\dagger_{\vec{k}^{'},\sigma'},a^\dagger_{\vec{k},\sigma}\big>=0
\end{eqnarray}
valid in all orders of the perturbation theory. The statistical
average values valid in the first order of the perturbation theory
have the following forms
\begin{eqnarray}\label{4.12}
\big<a^\dagger_{\vec{k}^{'},\sigma'},a_{\vec{k},\sigma}\big>_{01}=
\frac{\delta_{\vec{k},\vec{k}^{'}}\delta_{\sigma,\sigma'}}
{e^{\beta \xi_{\vec{k}}}+1}
\end{eqnarray}
\begin{eqnarray}\label{4.13}
\nonumber \big<H_I\big>_{01}=
-\frac{g}{V}\sum_{\vec{k}}\frac{\Theta(\hbar \omega_D-\mid
\xi_{\vec{k}} \mid)}{(e^{\beta \xi_{\vec{k}}}+1)^2}=\\=-gN(0)k_B T
\bigg( \ln \bigg[\frac{1+\tanh \frac{\beta \hbar
\omega_D}{2}}{1-\tanh \frac{\beta \hbar \omega_D}{2}}\bigg] -\tanh
\frac{\beta \hbar \omega_D}{2} \bigg)\;,
\end{eqnarray}
where
\begin{eqnarray}\label{4.14}
N(0)=\frac{mk_F}{2\pi^2\hbar^2}
\end{eqnarray}
is the density of electron states for one spin projection at the
Fermi surface. In evaluating $\big<H_I\big>_{01}$ the relations
$\mu \approx \epsilon_F$ and $\hbar \omega_D \ll \epsilon_F$ have
been used. From the relation (\ref{4.13}) one sees that
$\big<H_I\big>_{01}$ does not scale with the volume $V$ and the
same is true for all remaining terms $R(g)$ in (\ref{4.7}), as was
formally proven in \cite{bog} and demonstrated by explicit
calculations in \cite{kal}. The result (\ref{4.13}) together with
(\ref{4.7}) implies the relation
\begin{eqnarray}\label{4.15}
\Omega-\Omega_n=\big<H_I\big>_{01}+R(g) \;.
\end{eqnarray}
The right-hand side of the last equation is not proportional to
the volume $V$ and by this fact the density of the grand canonical
potential $\frac{\Omega}{V}$ approaches the density  
$\frac{\Omega_n}{V}$ corresponding to the ideal electron gas 
in the thermodynamic limit $V \to
\infty$. In other words, the interaction term $H_I$ in (\ref{4.1})
has no macroscopic effects on $\Omega(T,V,\mu;g)$ in the
thermodynamic limit $V \to \infty$ provided that one has chosen
the unperturbed Hamiltonian $H_0=H_{01}$ as given by (\ref{4.8}),
i.e.,
\begin{eqnarray}\label{4.16}
\Omega(T,V,\mu;g)=\Omega_n(T,V,\mu)
\end{eqnarray}
for the chosen inequivalent representation in the thermodynamic
limit.

Next we choose the unperturbed Hamiltonian $H_{02}$, to be the
one corresponding to the standard BCS theory of superconductivity,
i.e. in the form
\begin{eqnarray}\label{4.17}
H_{02} = \frac{V}{g}\Delta^{\ast}\Delta + \sum_{\vec{k},\sigma}
\xi_{\vec{k}}a^{\dag}_{\vec{k},\sigma}a_{\vec{k},\sigma} -
\sum_{\vec{k}} \bigg(\Delta
a^{\dagger}_{\vec{k},+}a^{\dagger}_{-\vec{k},-} + \Delta^{\ast}
a_{-\vec{k},-}a_{\vec{k},+} \bigg)\Theta(\hbar\omega_{D} - \mid
\xi_{{\vec k}}\mid) \;,
\end{eqnarray}
where $\Delta$ and $\Delta^{\ast}$ are the following average
values
\begin{eqnarray}\label{4.18}
\nonumber \Delta = \frac{g}{V} \sum_{\vec{k}}
\big<a_{-\vec{k},-}a_{\vec{k},+}\big>_{02} \Theta(\hbar\omega_{D}
- \mid
\xi_{{\vec k}}\mid) \;, \\
\Delta^{\ast} = \frac{g}{V} \sum_{\vec{k}}
\big<a^{\dagger}_{\vec{k},+}a^{\dagger}_{-\vec{k},-}\big>_{02}
\Theta(\hbar\omega_{D} - \mid \xi_{{\vec k}}\mid)
\end{eqnarray}
called the gap functions. $H_{02}$ is the Hamiltonian (\ref{4.1})
in the so called mean field approximation.

\begin{figure}[h]
\bigskip{}
\centering\resizebox{0.6\textwidth}{!}{\includegraphics{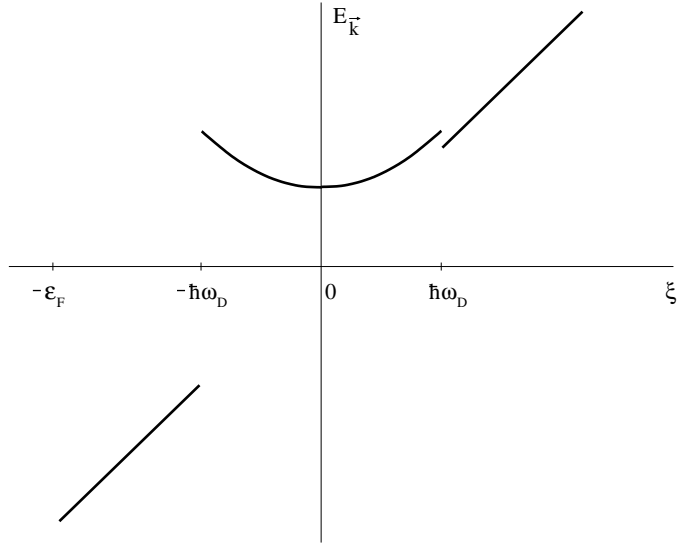}
\par} \caption{The energy spectrum $E_{\vec{k}}$ as function of $\xi$.}
\label{fig1}
\bigskip{}
\end{figure}

The unperturbed Hamiltonian $H_{02}$ is diagonalized by means of
transformations (\ref{2.8}), (\ref{2.9}) with the transformation
parameters $\alpha_{\vec{k}}$ determined by the formula
\begin{eqnarray}\label{4.19}
\sin^2\alpha_{\vec{k}}=\frac{1}{2}\Bigg( 1-\frac{\xi_{\vec{k}}}
{\sqrt{\xi^2_{\vec{k}}+ \Delta^2}} \Bigg) \Theta(\hbar\omega_D -
\mid \xi_{\vec{k}} \mid) \;\;, \Delta=\Delta^{\ast}\;.
\end{eqnarray}
All the transformation parameters $\alpha_{\vec{k}}$ are
determined in terms of the physical parameters entering $H_{02}$
and specify the single inequivalent representation of the
anticommutator ring (\ref{2.15}) with the auxiliary condition
(\ref{2.16}). Its representation space is the Hilbert space
denoted by $\mathcal{H}_2$. The unperturbed Hamiltonian $H_{02}$
has the following diagonal form
\begin{eqnarray}\label{4.20}
\widetilde{H}_{02}=\frac{V}{g}\Delta^2+\sum_{\vec{k},\sigma}
\bigg[E_{\vec{k}}c^{\dagger}_{\vec{k},\sigma}c_{\vec{k},\sigma}+
\frac{1}{2}\big(\xi_{\vec{k}}-E_{\vec{k}}\big)\Theta(\hbar
\omega_D-\mid \xi_{\vec{k}} \mid) \bigg]\;,
\end{eqnarray}
where $E_{\vec{k}}$ is the energy spectrum of elementary
excitations given by the formulae
\begin{eqnarray}\label{4.21}
\nonumber E_{\vec{k}}=\sqrt{\Delta^2+\xi^2_{\vec{k}}}
\;\Theta(\hbar \omega_D-\mid
\xi_{\vec{k}} \mid)\;,\\
E_{\vec{k}}=\xi_{\vec{k}}\;\Theta(\mid \xi_{\vec{k}} \mid-\hbar
\omega_D)\;\; ; \;\; \xi_{\vec{k}} \in <-\mu,+\infty> \;.
\end{eqnarray}
The energy spectrum $E_{\vec{k}}$ as the function of $\xi$ is
schematically depicted on Fig. \ref{fig1}.

As is seen from the Fig. \ref{fig1} the energy spectrum
$E_{\vec{k}}$ has discontinuities at two fixed points $\xi=\pm
\hbar \omega_D$ which are specified by the material parameter
$\omega_D$.

The unperturbed Hamiltonian $H_{02}$ as given by (\ref{4.17}) or
(\ref{4.20}) both determines and is determined by the average
values (\ref{4.18}). Thus the relation (\ref{4.18}) is, in fact,
the self-consistency condition known in all text books
\cite{fetter} as the gap equation
\begin{eqnarray}\label{4.22}
\Delta=g\frac{\Delta}{(2\pi)^3}\int\frac{d^3\vec{k}}{2E_{\vec{k}}}\tanh
\frac{1}{2}\beta E_{\vec{k}}\;\Theta(\hbar \omega_D-\mid
\xi_{\vec{k}} \mid)\;.
\end{eqnarray}
The gap function $\Delta$ is different from zero, $\Delta \neq 0$,
for $T<T_c$, where $T_c$ is the critical temperature given by the
formula \cite{fetter}
\begin{eqnarray}\label{4.23}
T_c=\frac{2e^{\gamma}}{\pi}T_D \exp{\bigg[-\frac{1}{gN(0)}}\bigg]
\doteq 1.13\;T_D \exp{\bigg[-\frac{1}{gN(0)}}\bigg] \;.
\end{eqnarray}
Here $\gamma \doteq 1.781$ is Euler constant and $T_D=\frac{\hbar
\omega_D}{k_B}$ is the Debye temperature. The critical temperature
$T_c$, as seen from (\ref{4.23}), is proportional to $T_D$ and is
a singular function of $g$ at $g=0$.

The solution to the gap equation (\ref{4.22}) provides
$\Delta=\Delta(T,g)$ as a function of $T$ and $g$, which cannot be
expressed in an analytic form, but only numerically. The only
contribution to $\big<H_I\big>(g)$ which survives the
thermodynamic limit $V \to \infty$ has the form
\begin{eqnarray}\label{4.24}
\big<H_I\big>(g)=\big<H_I\big>_{02}(g)=-\frac{V}{g}\Delta^2(g)
\end{eqnarray}
because all remaining terms denoted by $R(g)$ in (\ref{4.7})
become negligible in this limit as formally proven in \cite{bog}
and demonstrated by explicit calculations in \cite{kal}.

Two different results (\ref{4.13}) and (\ref{4.24}) for the
average value of the same physical abservable $\big<H_I\big>$
exhibit clearly and evidently the important role of the
inequivalent representations in practical applications. Namely,
$\big<H_I\big>$ has no macroscopic effects if evaluated in
the representation Hilbert space $\mathcal{H}_1$, however,
$\big<H_I\big>$ has a very relevant macroscopic contribution if
evaluated in the representation Hilbert space $\mathcal{H}_2$.

The result (\ref{4.24}) with the relation (\ref{4.5}) gives the
expression for the grand canonical potential $\Omega(T,V,\mu;g)$
in the form
\begin{eqnarray}\label{4.25}
\Omega(T,V,\mu;g)=\Omega_n(T,V,\mu)-\frac{V}{8\pi}H^2_c(T)\;,
\end{eqnarray}
where $H_c(T)$ is the thermodynamic critical magnetic field
defined by the relation
\begin{eqnarray}\label{4.26}
H^2_c(T)=8\pi \int_0^g \frac{dg'}{g'^2}\;\Delta^2(g')\;.
\end{eqnarray}
The solution to the BCS Hamiltonian (\ref{4.1}), as represented by
the relations (\ref{4.20})-(\ref{4.26}), describes, as is
well-known, the superconducting state of the system governed by
the Hamiltonian (\ref{4.1}). This solution will be referred as to
the standard solution of the BCS theory of superconductivity. We
have discussed it briefly in order to demonstrate the role of
inequivalent representations of the canonical anticommutator ring
(\ref{2.1}) or (\ref{2.15}) of electron field operators on
solutions which are generally very well-known.

The main purpose of this section is to explore the physical
implications of an additional inequivalent representation of the
anticommutator ring (\ref{2.15}) associated with the Hamiltonian
(\ref{4.1}) of the BCS theory. Without any physical motivation we
choose the third form of the unperturbed Hamiltonian $H_{03}$, as
given by
\begin{eqnarray}\label{4.27}
\nonumber H_{03} = \frac{g}{V}\sum_{\vec{k},\vec{k}^{'}}\big<
a^{\dagger}_{\vec{k},+}a^{\dagger}_{-\vec{k},-}\big>_{03}\big<
a_{-\vec{k}^{'},-}a_{\vec{k}^{'},+}\big>_{03}\Theta(\hbar\omega_{D}
- \mid \xi_{{\vec k}}\mid)\;\Theta(\hbar\omega_{D} - \mid
\xi_{{\vec k}^{'}}\mid)\;\Theta(q-\mid k-k^{'}\mid)+\\+
\sum_{\vec{k},\sigma}
\xi_{\vec{k}}a^{\dag}_{\vec{k},\sigma}a_{\vec{k},\sigma} -
\sum_{\vec{k}} \bigg(\Delta_{\vec{k}}
a^{\dagger}_{\vec{k},+}a^{\dagger}_{-\vec{k},-} +
\Delta^{\ast}_{\vec{k}} a_{-\vec{k},-}a_{\vec{k},+}
\bigg)\Theta(\hbar\omega_{D} - \mid \xi_{{\vec k}}\mid) \;,\quad
\end{eqnarray}
where
\begin{eqnarray}\label{4.28}
\nonumber \Delta_{\vec{k}} = \frac{g}{V} \sum_{\vec{k}^{'}}
\big<a_{-\vec{k}^{'},-}a_{\vec{k}^{'},+}\big>_{03}\Theta(\hbar\omega_{D}
- \mid
\xi_{{\vec{k}^{'}}}\mid)\;\Theta(q-\mid k-k^{'}\mid) \;, \\
\Delta^{\ast}_{\vec{k}} = \frac{g}{V} \sum_{\vec{k}^{'}}
\big<a^{\dagger}_{\vec{k}^{'},+}a^{\dagger}_{-\vec{k}^{'},-}\big>_{03}
\Theta(\hbar\omega_{D} - \mid
\xi_{{\vec{k}^{'}}}\mid)\;\Theta(q-\mid k-k^{'}\mid)
\end{eqnarray}
are gap functions which are dependent on the wave vector
$\vec{k}$. Here, $q$ is chosen to be the absolute value of the
wave vector corresponding to the minimal energy of an electron
confined in a box with the edges $L_1,L_2$ and $L_3$, i.e.,
\begin{eqnarray}\label{2.34a}
\frac{\hbar^2q^2}{2m}=\frac{\hbar^2}{2m}\bigg[
\bigg(\frac{\pi}{L_1}\bigg)^2+\bigg(\frac{\pi}{L_2}\bigg)^2+
\bigg(\frac{\pi}{L_3}\bigg)^2\bigg] \;.
\end{eqnarray}
One may say that $H_{03}$ is a kind of an interpolation between
the Hamiltonians $H_{01}$ and $H_{02}$. Indeed, $H_{03}$ contains
only those terms from $H_{02}$ which partially conserve the energy
in microscopic scattering processes between electrons. This fact
is represented by the presence of the step function
$\Theta({q-\mid k-k^{'}\mid})$ in (\ref{4.27}) and (\ref{4.28}).
One may regard $H_{03}$ as a mathematical toy in order to search
for an additional inequivalent representation of the
anticommutator ring (\ref{2.15}) within the framework of the same
Hamiltonian (\ref{4.1}).

The Hamiltonian $H_{03}$ is again diagonalized by the well-known
Bogoliubov-Valatin transformations (\ref{2.8})-(\ref{2.9}). The
parameters $\alpha_{\vec{k}}$ of the transformations
(\ref{2.4})-(\ref{2.9}) are defined by the relation
\begin{eqnarray}\label{4.30}
\sin\alpha_{\vec{k}}=-\frac{1}{\sqrt{2} }\Bigg(
1-\frac{\xi_{\vec{k}}} {\sqrt{\xi^2_{\vec{k}}+\mid
\Delta_{\vec{k}} \mid^2}} \Bigg)^{\frac{1}{2}}
\Theta(\hbar\omega_D - \mid \xi_{\vec{k}} \mid)
\end{eqnarray}
with the gap functions $\Delta_{\vec{k}}$ depending on
$\vec{k}$, which specify the chosen inequivalent representation of
the anticommutator ring (\ref{2.15}). In this representation 
the unperturbed Hamiltonian $H_{03}$ gets its diagonal form
\begin{eqnarray}\label{4.31}
\nonumber \widetilde{H}_{03}=\frac{g}{V}\sum_{\vec{k},\vec{k}^{'}}
\big< a^{\dagger}_{\vec{k},+}a^{\dagger}_{-\vec{k},-}\big>_{03}
\big<a_{-\vec{k}^{'},-}a_{\vec{k}^{'},+}\big>_{03}
\Theta(\hbar\omega_{D} - \mid
\xi_{{\vec{k}}}\mid)\;\Theta(\hbar\omega_{D} - \mid
\xi_{{\vec{k}^{'}}}\mid)\;\Theta(q-\mid
k-k^{'}\mid)+\\+\sum_{\vec{k},\sigma}
\bigg[E_{\vec{k}}c^{\dagger}_{\vec{k},\sigma}c_{\vec{k},\sigma}+
\frac{1}{2}\big(\xi_{\vec{k}}-E_{\vec{k}}\big)\Theta(\hbar
\omega_D-\mid \xi_{\vec{k}} \mid) \bigg]\;,\quad
\end{eqnarray}
where
\begin{eqnarray}\label{4.32}
\nonumber E_{\vec{k}}=\sqrt{\mid \Delta_{\vec{k}}
\mid^2+\xi^2_{\vec{k}}} \;\Theta(\hbar \omega_D-\mid
\xi_{\vec{k}} \mid)\;,\\
E_{\vec{k}}=\xi_{\vec{k}}\;\Theta(\mid \xi_{\vec{k}} \mid-\hbar
\omega_D)\;\; ; \; \xi_{\vec{k}} \in (-\mu,+\infty)
\end{eqnarray}
is the energy spectrum of quasiparticles represented by the field
operators $c_{\vec{k},\sigma}$ and $c^{\dagger}_{\vec{k},\sigma}$.
By evaluating the statistical average value
$\big<a_{-\vec{k},-}a_{\vec{k},+}\big>_{03}$ we get the following
relation
\begin{eqnarray}\label{4.33}
\big<a_{-\vec{k},-}a_{\vec{k},+}\big>_{03}=\frac{1}{2E_{\vec{k}}}
\Delta_{\vec{k}} \tanh \frac{\beta E_{\vec{k}}}{2}\Theta(\hbar
\omega_D-\mid \xi_{\vec{k}} \mid)\;.
\end{eqnarray}
The last relation, when inserted into the definitions
(\ref{4.28}), gives us the self-consistency condition for the gap
function
\begin{eqnarray}\label{4.34}
\Delta_{\vec{k}}=\frac{g}{V}\sum_{\vec{k}^{'}}
\frac{\Delta_{\vec{k}^{'}}}{2E_{\vec{k}^{'}}}\tanh \frac{\beta
E_{\vec{k}^{'}}}{2}\Theta(\hbar \omega_D-\mid
\xi_{\vec{k}^{'}}\mid) \;\Theta(q-\mid k-k^{'}\mid)\;.
\end{eqnarray}
The equations (\ref{4.33}) and (\ref{4.34}) determining
$\big<a_{-\vec{k},-}a_{\vec{k},+}\big>_{03}$ and the gap function
$\Delta_{\vec{k}}$ resp. are, of course, closely related to those
ones in the standard solution to the BCS theory of the
superconductivity except for the $\vec{k}$ dependence of
$\Delta_{\vec{k}}$ and the presence of the step function
$\Theta(q-\mid k-k^{'}\mid)$ in (\ref{4.34}) by which a partial
energy conservation rule is required even for microscopic
scattering processes between electrons.

\newpage
\section{The solution to the gap equation}\label{sec5}
The solution to the gap equation (\ref{4.34}) is analyzed
similarly as in the work \cite{cron}. The assumption that the gap
function $\Delta_{\vec{k}}$ is a function of the magnitude $k=\mid
\vec{k} \mid$ only, simplifies the relation (\ref{4.34})
substantially. We employ the inequality $q\ll k^{'}$ which is
valid for every vector $\vec{k}^{'}$ in the sum (\ref{4.34}). By
this fact, we can replace the equation (\ref{4.34}) by the
relation
\begin{eqnarray}\label{5.1}
\Delta_{\vec{k}}=\frac{g}{2V}\frac{\nu(k)}{E_{\vec{k}}}\Delta_{\vec{k}}
\tanh \frac{\beta E_{\vec{k}}}{2}\;\Theta(\hbar \omega_D-\mid
\xi_{\vec{k}}\mid)\;
\end{eqnarray}
where
\begin{eqnarray}\label{5.2}
\nu(k)=\sum_{\vec{k}^{'}}\Theta(q-\mid k-k^{'}\mid)\;\Theta(\hbar
\omega_D-\mid \xi_{\vec{k}^{'}} \mid)\;.
\end{eqnarray}
We emphasize that the equation (\ref{5.1}) takes into account only
linear terms in the tiny quantity $q$. The quantity $\nu(k)$, as
given by the equation (\ref{5.2}), is, in fact, the number of wave
vectors $\vec{k}^{'}$ with the lengths $k^{'}$ from the interval
$k-q<k^{'}<k+q$. In order to evaluate the number $\nu(k)$ we
remind ourselves that the vector $\vec{k}^{'}$ have the components
given below
\begin{eqnarray}\label{5.3}
\vec{k}^{'}=\bigg(\frac{2\pi}{L_1}n_1,\frac{2\pi}{L_2}n_2,
\frac{2\pi}{L_3}n_3 \bigg)\;,
\end{eqnarray}
where $n_1$, $n_2$, $n_3$ are integers and $L_1$, $L_2$, $L_3$ are
edge lengths of a box into which the electron system
is enclosed. The given magnitude $k^{'}$ defines the surface of an
ellipsoid with semiaxes $a_1=k^{'}\frac{L_1}{2\pi}$,
$a_2=k^{'}\frac{L_2}{2\pi}$ and $a_3=k^{'}\frac{L_3}{2\pi}$ in an
Euclidean space with coordinates $n_1,n_2$ and $n_3$, i.e.,
\begin{eqnarray}\label{5.4}
\frac{n_1^2}{a_1^2}+\frac{n_2^2}{a_2^2}+\frac{n_3^2}{a_3^2}=1\;.
\end{eqnarray}
Thus the number of the states $\nu(k)$ is approximately equal to
the volume enclosed between two surface areas of the ellipsoids
corresponding to the minimal and maximal semiaxes $a_1, a_2$ and
$a_3$, i.e.,
\begin{eqnarray}\label{5.5}
\nu(k)=\frac{1}{\pi^2}\;qL_1L_2L_3k^2\;\Theta(\hbar \omega_D-\mid
\xi_{\vec{k}}\mid)\;.
\end{eqnarray}
We analyze the number $\nu(k)$ for two limiting cases, namely,
when the ellipsoid (\ref{5.4}) degenerates either into a sphere or
into a thin circular disc. The sphere corresponds to the case when
the electron system under consideration is enclosed in a cube
with the edges $L_1=L_2=L_3=L$, i.e., it has isotropic bulk
properties. If the ellipsoid (\ref{5.4}) degenerates into a thin
circular disc with $L_1=L_2\gg L_3=d$, then the system of
electrons has a form of a thin film of the given thickness $d$,
i.e., it has anisotropic properties. Both of these limiting cases
represent objects of the physical interest.

We first analyze the case corresponding to the isotropic bulk
material. In this case we get
\begin{eqnarray}\label{5.6}
\nu(k)= \frac{2\sqrt{3}}{\pi}\;k^2L^2\;\Theta(\hbar \omega_D-\mid
\xi_{\vec{k}}\mid)\;.
\end{eqnarray}
By inserting the last result with $V=L^3$ into the equation
(\ref{5.1}) we get the relation
\begin{eqnarray}\label{5.7}
\Delta_{\vec{k}}=\frac{\sqrt{3}g}{\pi
L}\;\Delta_{\vec{k}}\;\frac{k^2}{E_{\vec{k}}}\;\tanh \frac{\beta
E_{\vec{k}}}{2}\;\Theta(\hbar \omega_D-\mid \xi_{\vec{k}}\mid)\;
\end{eqnarray}
which in the thermodynamic limit $L \to \infty$ has the only
solution $\Delta_{\vec{k}}=0$. Thus this type of superconductivity
cannot exist in three dimensional isotropic materials.

We next analyze the second limiting case when the system has a
form of a thin film with a given thickness $d$. In this case
$V=L^2d$, $q\doteq \frac{\pi}{d}$ and the number of the states
$\nu(k)$ is given by the relation
\begin{eqnarray}\label{5.8}
\nu(k)= \frac{k^2L^2}{\pi}\;\Theta(\hbar \omega_D-\mid
\xi_{\vec{k}}\mid)\;.
\end{eqnarray}
By inserting the last result into equation (\ref{5.1}) we get the
relation
\begin{eqnarray}\label{5.9}
\Delta_{\vec{k}}=\frac{g}{2\pi
d}\frac{k^2}{E_{\vec{k}}}\;\Delta_{\vec{k}} \tanh \frac{\beta
E_{\vec{k}}}{2}\;\Theta(\hbar \omega_D-\mid \xi_{\vec{k}}\mid)\;
\end{eqnarray}
which can have a nontrivial solution $\Delta_{\vec{k}}\neq 0$ in
the thermodynamic limit $L \to \infty$.

Since the vectors $\vec{k}$ in the relation (\ref{4.28}) are
restricted by the inequalities below
\begin{eqnarray}\label{5.10}
\epsilon_F-\hbar \omega_D \leq \frac{\hbar^2k^2}{2m}\leq
\epsilon_F+\hbar \omega_D\;,\;\hbar \omega_D \ll \epsilon_F\;
\end{eqnarray}
we can replace the quantity $k^2$ in (\ref{5.9}) by $k_F^2$. Then
from the equation (\ref{5.9}) it follows that the energy spectrum
$E_{\vec{k}}$ of the quasiparticles in the superconducting state
with $\Delta_{\vec{k}}\neq 0$ must satisfy the following relation
\begin{eqnarray}\label{5.11}
1=G\frac{2\epsilon_F}{E_{\vec{k}}}\;\tanh
\frac{\beta E_{\vec{k}}}{2}\;\Theta(\hbar \omega_D-\mid
\xi_{\vec{k}}\mid)\;
\end{eqnarray}
where
\begin{eqnarray}\label{5.12}
G=\frac{g}{4\pi^2}\bigg(\frac{2m}{\hbar^2}\bigg)^{\frac{3}{2}}
\sqrt{\epsilon_1}
\end{eqnarray}
is the dimensionless effective coupling constant and
\begin{eqnarray}\label{5.13}
\epsilon_1=\frac{\hbar^2}{2m}\frac{\pi^2}{d^2}
\end{eqnarray}
is the minimal energy of an electron confined in a thin layer of
the thickness $d$.

The equation (\ref{5.11}) determines the quasiparticle energy spectrum
$E_{\vec{k}}$ as a certain function of
temperature $T$. It can be satisfied only if the temperature $T$
is below the critical temperature $T_c$,
\begin{eqnarray}\label{5.14}
T_c=GT_F\;,
\end{eqnarray}
where $T_F$ is the Fermi temperature. It is interesting to point
out that the relation (\ref{5.14}) between the critical
temperature $T_c$ and the Fermi temperature $T_F$ has accidently
the same form as the one found from the analysis of experimental data
by Uemura et al. \cite{uemur} and discussed in \cite{pist} on a
phenomenological basis. In the phenomenological formula of the
form (\ref{5.14}), proposed in \cite{uemur} and called as the
Uemura plot in \cite{pist}, the phenomenological parameter $G$ is
the same for a unique group of HTS. The universal correlation
between $T_c$ and $T_F$ as given by (\ref{5.14}) has been claimed
\cite{uemur,pist} to exist in all HTS with planar structures. The
coincidence between the theoretical formula (\ref{5.14}) and the
phenomenological formula \cite{uemur,pist} seems to be promising
and stimulates us for further theoretical investigation of this
type of superconductivity to be compared with the experimental
data of planar HTS.

The relation (\ref{5.14}) determining the critical temperature
$T_c$ differs qualitatively from the relation (\ref{4.23})
corresponding to the standard solution of the BCS theory. $T_c$ is
proportional to $T_F$ in contrast to the Debye temperature $T_D$
in (\ref{4.23}), i.e., $T_c$ does not depend on ionic mass. Thus
this type of superconductivity, despite of the fact that it is due
to electron-phonon interaction, does not exhibit the isotope
effect. $T_c$ is an analytic function of $g$ at $g=0$. The
proportionality coefficient $G$ depends on the lowest bound energy
$\epsilon_1$ for the energy spectrum of electrons confined in a
layer of the thickness $d$, i.e., for the thinner layer one has
the higher critical temperature $T_c$. For an isotropic bulk
material $\epsilon_1 \to 0$ and in the same way $T_c \to 0$. This
property seems to represent the reason why this type of
superconductivity cannot exist in isotropic systems, but it can
appear in systems with electrons residing on quasi two dimensional
planar structures. All these features seem to be promising that
this type of superconductivity may have some relevance for the
properties of HTS observed in experiments.

\begin{figure}[h]
\bigskip{}
\centering\resizebox{0.6\textwidth}{!}{\includegraphics{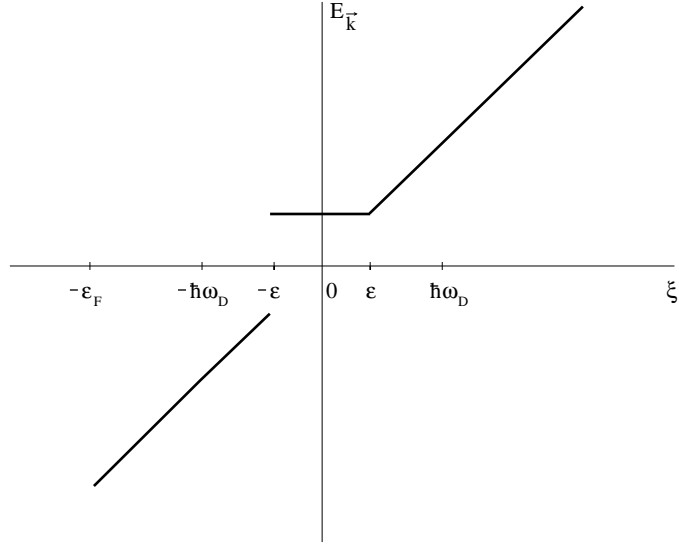}
\par} \caption{The energy spectrum $E_{\vec{k}}$ of quasiparticles
as a function of $\xi$.} \label{fig2}
\bigskip{}
\end{figure}

The solution to equation (\ref{5.11}) for the energy spectrum
$E_{\vec{k}}$, in the range of $\vec{k}$ for which
$\Delta_{\vec{k}}\neq 0$, is a quantity $\epsilon(T)$ which is
independent of $\vec{k}$. However, it is a function of $T$, i.e.,
\begin{eqnarray}\label{5.15}
\nonumber E_{\vec{k}}=\sqrt{\mid \Delta_{\vec{k}}
\mid^2+\xi^2_{\vec{k}}}=\epsilon(T) \;,\;T\leq T_c\;,\;\mid
\xi_{\vec{k}} \mid)\leq \epsilon(T)\leq \hbar \omega_D\;\;,\\
E_{\vec{k}}=\xi_{\vec{k}}\;,\;\mid \xi_{\vec{k}} \mid \geq
\epsilon(T)\;,\;\xi_{\vec{k}} \in (-\mu,+\infty) \;.
\end{eqnarray}
One cannot obtain an explicit expression for the solution
$\epsilon(T)$ to the equation (\ref{5.11}) in the form of
elementary functions, but has to resort to numerical methods given
below. The energy spectrum $E_{\vec{k}}$ as a function of $\xi$ is
depicted on the Fig. \ref{fig2}.

The energy spectrum $E_{\vec{k}}$ as given by the formulae
(\ref{5.15}) differs quantitatively from that one given by the
relation (\ref{4.21}) and depicted on the Fig. \ref{fig1}
corresponding to the standard solution of the BCS theory. The
energy spectrum (\ref{5.15}) has no discontinuities at fixed
points, at $\mid \xi \mid=\hbar \omega_D$, in contrast to the
standard solution of the BCS theory. It has a single
discontinuity, at the point $\xi=-\epsilon$, which is a function
of the temperature $T$.

The function $\epsilon(T)$ determined by the transcendental
equation
\begin{eqnarray}\label{5.16}
1=G\frac{2\epsilon_F}{\epsilon}\tanh \frac{\beta \epsilon}{2}
\end{eqnarray}
has the following behavior for the limiting cases
\begin{eqnarray}\label{5.17}
\epsilon(T)=\epsilon_0\big(1-2e^{-\beta \epsilon_0}\big) \;,\; T
\ll T_c
\end{eqnarray}
and
\begin{eqnarray}\label{5.18}
\epsilon(T)=\sqrt3\;
\epsilon_0\bigg(1-\frac{T}{T_c}\bigg)^{\frac{1}{2}} \;,\; T_c-T
\ll T_c\;,
\end{eqnarray}
where
\begin{eqnarray}\label{5.19}
\epsilon_0\equiv \epsilon(T=0)=2G\epsilon_F=2k_BT_c\;.
\end{eqnarray}
Thus the ratio $\frac{\epsilon_0}{k_BT_c}=2$ is a universal
constant, independent on material parameters of the thin layer
under consideration.

\begin{figure}[h]
\bigskip{}
\centering\resizebox{0.6\textwidth}{!}{\includegraphics{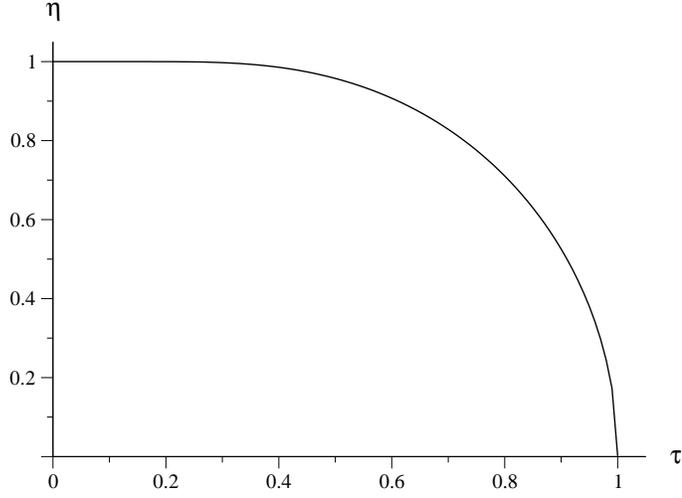}
\par} \caption{The energy spectrum $\epsilon(T)=\eta(\tau)\epsilon_0$
as a function of the reduced temperature $\tau$.} \label{fig3}
\bigskip{}
\end{figure}

For the numerical analysis of the solution $\epsilon(T)$ to the
equation (\ref{5.16}) it is convenient to introduce the following
dimensionless variables
\begin{eqnarray}\label{5.20}
\eta(T)=\frac{\epsilon(T)}{\epsilon_0} \;,\; \tau=\frac{T}{T_c}\;,
\end{eqnarray}
where $\tau$ is the so called reduced temperature. In this
notation the function $\eta(T)$ is expressed in the implicit form
\begin{eqnarray}\label{5.21}
\eta=\tanh \bigg( \frac{\eta}{\tau} \bigg)
\end{eqnarray}
and its numerical representation is shown on the Fig. \ref{fig3}.

From the relations (\ref{4.33}) and (\ref{5.15}) we get the
explicit expression for the following average value
\begin{eqnarray}\label{5.22}
\big<a_{-\vec{k},-}a_{\vec{k},+} \big>_{03}=
\frac{1}{4G\epsilon_F}(\epsilon^2-\xi^2_{\vec{k}})^{\frac{1}{2}}
\Theta(\epsilon-\mid \xi_{\vec{k}}
\mid)\;\Theta\bigg(G-\frac{1}{\beta \epsilon_F}\bigg)\;,\;
\epsilon<\hbar \omega_D \;,
\end{eqnarray}
which is of great importance for the calculation of
thermodynamic properties of the novel superconducting
state.

\newpage
\section{Thermodynamical properties of the superconducting state}\label{sec6}
All thermodynamic properties of the superconducting state are
extracted from the difference (\ref{4.5}) between the grand
canonical potential $\Omega(T,V,\mu;g) \equiv \Omega_s$
corresponding to the superconducting state for $T<T_c$ and
$\Omega(T,V,\mu;g=0) \equiv \Omega_n$ corresponding to the normal
state. For this reason we first evaluate the average value
$\big<H_I\big>(g)$. By using the result (\ref{5.22}) we get the
expression
\begin{eqnarray}\label{6.1}
\langle H_I \rangle(g) = -\frac{g}{V}\;
\frac{\Theta\big(G-\frac{1}{\beta
\epsilon_F}\big)}{16G^2\epsilon^2_F} \sum_{\vec{k},
\vec{k}^{'}}\,(\epsilon^2-\xi^2_{\vec{k}})^{\frac{1}{2}}
(\epsilon^2-\xi^2_{\vec{k}^{'}})^{\frac{1}{2}}\; \Theta(\epsilon -
\mid \xi_{{\vec{k}}}\mid )\; \Theta(\epsilon - \mid
\xi_{{\vec{k^{'}}}}\mid ) +R(g)\;.
\end{eqnarray}
The above sum over $\vec{k}$ and $\vec{k}^{'}$ is evaluated as an
integral by the standard substitution
\begin{eqnarray}\label{6.1a}
\nonumber \sum_{\vec{k},\vec{k}^{'}}\dots \to \frac{V^2}{(2\pi)^6}
\int d^3\vec{k} d^3\vec{k}^{'}
\end{eqnarray}
and by using the constraints (\ref{5.10}) we get the result
\begin{eqnarray}\label{6.2}
\langle H_I \rangle(g) =
-\frac{1}{(2\pi)^2}\frac{gV}{2^5G^2\epsilon_F} \bigg(
\frac{m}{\hbar^2} \bigg)^3
\epsilon^4(T)\Theta\bigg(G-\frac{1}{\beta \epsilon_F}\bigg)+R(g)
\;,
\end{eqnarray}
where $R(g)$ denotes all remaining terms in the perturbation
series which are negligible in the thermodynamic limit $V \to
\infty$ similarly as for the perturbation series with unperturbed
Hamiltonians $H_{01}$ and $H_{02}$.

Now, we are ready to express the difference (\ref{4.5}) as the
following integral
\begin{eqnarray}\label{6.3}
\Omega_s-\Omega_n=-\frac{Vg}{2^5k_BT_c}\frac{1}{(2\pi)^2}\bigg(
\frac{m}{\hbar^2} \bigg)^3 \int_{\frac{1}{\beta \epsilon_F}}^G
\frac{dG'}{G'^2} \; \epsilon^4 \;,
\end{eqnarray}
where $G$ is the dimensionless coupling constant defined by
equation (\ref{5.12}). The last integral is luckily in a
particularly convenient form because the relation (\ref{5.16})
expresses $G$ as the following function of $\epsilon$,
\begin{eqnarray}\label{6.4}
\frac{1}{G}=\frac{2\epsilon_F}{\epsilon}\tanh \frac{\beta
\epsilon}{2}
\end{eqnarray}
The direct substitution of the last relation into the integral
(\ref{6.3}) leads to the formula
\begin{eqnarray}\label{6.5}
\Omega_s-\Omega_n=\frac{V}{(2\pi^2)}\frac{g\epsilon_F}{16k_BT_c}\bigg(
\frac{m}{\hbar^2} \bigg)^3 \int_0^{\epsilon}d\epsilon' \epsilon'^4
\frac{d}{d\epsilon'} \bigg( \frac{1}{\epsilon'}\tanh \frac{\beta
\epsilon'}{2} \bigg) \;.
\end{eqnarray}
Two per partes integrations of the last integral and the use of
the reduced variables $\eta=\frac{\epsilon}{\epsilon_0}$ and
$\tau=\frac{T}{T_c}$ give us the result
\begin{eqnarray}\label{6.6}
\Omega_s-\Omega_n=-\frac{V}{6(2\pi)^2}\;g\epsilon_F(k_BT_c)^2\bigg(
\frac{m}{\hbar^2} \bigg)^3
\bigg[\eta^4-\frac{1}{2}\tau^3\varphi(\eta)\bigg]\;,
\end{eqnarray}
where $\varphi(\eta)$ is a function of $\eta$ defined by the
integral
\begin{eqnarray}\label{6.7}
\varphi(\eta)=\int_0^{\eta}dx \bigg(
\ln{\frac{1+x}{1-x}}\bigg)^3\;.
\end{eqnarray}

By passing from the grand canonical potential $\Omega(T,V,\mu)$ to
the Helmholtz free energy $F(T,V,N)$ we get the difference
\begin{eqnarray}\label{6.8}
\Omega_s-\Omega_n=F_s-F_n=-\frac{V}{8\pi}H_c^2(T) \;,
\end{eqnarray}
where $H_c(T)$ is the thermodynamic critical magnetic field given
by the formula
\begin{eqnarray}\label{6.9}
H_c(T)=H_c(0)\bigg[\eta^4-\frac{1}{2}\tau^3\varphi(\eta)\bigg]^{\frac{1}{2}}
\;.
\end{eqnarray}
Here, $H_c(0)$ is the critical magnetic field at $T=0$,
\begin{eqnarray}\label{6.10}
H_c(0)=\bigg( \frac{g\epsilon_F}{3\pi}\bigg)^{\frac{1}{2}}\bigg(
\frac{m}{\hbar^2}\bigg)^{\frac{3}{2}}=2\pi^2 \bigg(
\frac{g}{6\pi}\bigg)^{\frac{1}{2}}N(0)k_BT_c \;.
\end{eqnarray}

\begin{figure}[h]
\bigskip{}
\centering\resizebox{0.6\textwidth}{!}{\includegraphics{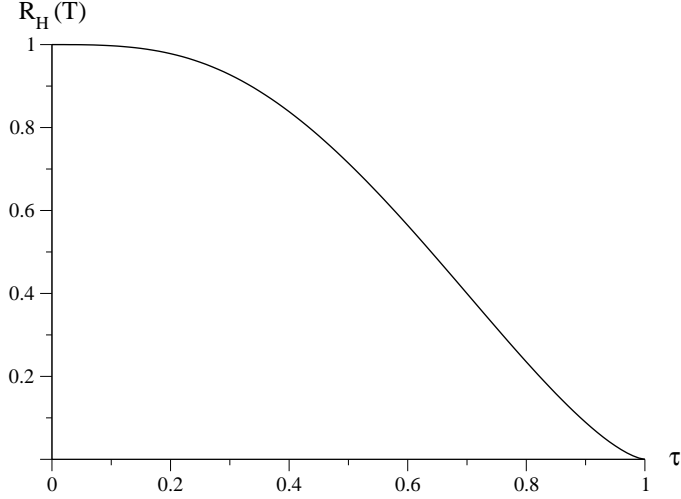}
\par} \caption{The temperature dependence of the critical magnetic field
$H_c(T)$ plotted as the ratio $R_H(\tau)=H_c(T)/H_c(0)$.}
\label{fig4}
\bigskip{}
\end{figure}

The temperature dependence of the ratio $R_H(\tau)=H_c(T)/H_c(0)$
is shown on the Fig. \ref{fig4}.
\\
The theoretical curve on the Fig. \ref{fig4} is very similar to
the experimental data of the upper critical magnetic field for HTS
$\mathrm{Bi_{2}Sr_{2}CaCu_{2}O_{8}}$ \cite{alex}. The behavior of
$H_c(T)$ at $T_c-T \ll T_c$ and $T \ll T_c$ is given by the
following formulae
\begin{eqnarray}\label{6.11}
H_c(T)= 3H_c(0)(1-\tau)^{\frac{3}{2}} \;,\; T_c-T \ll T_c
\end{eqnarray}
and
\begin{eqnarray}\label{6.12}
H_c(T)=H_c(0)\bigg( 1-\frac{9\zeta(3)}{4}\;\tau^3 \bigg) \;,\;T
\ll T_c \;,
\end{eqnarray}
where $\zeta(x)$ is the Riemann dzeta function, $\zeta(3)\approx
1.202$.

Now we compare the results (\ref{6.10})-(\ref{6.12}) with the
behavior of the critical magnetic field $H_c^{'}(T)$ corresponding
to the results of the standard solution of the BCS theory
\cite{fetter}
\begin{eqnarray}\label{6.13}
H_c^{'}(0)=\pi e^{-\gamma}[4\pi N(0)]^{\frac{1}{2}}k_BT^{'}_c
\doteq 1.76[4\pi N(0)]^{\frac{1}{2}}k_BT^{'}_c \;,
\end{eqnarray}
\begin{eqnarray}\label{6.14}
H_c^{'}(T)=
H_c^{'}(0)\;e^{\gamma}\bigg[\frac{8}{7\zeta(3)}\bigg]^{\frac{1}{2}}(1-\tau')\doteq
1.74H_c^{'}(0)(1-\tau')\;,\; T^{'}_c-T \ll T^{'}_c
\end{eqnarray}
and
\begin{eqnarray}\label{6.15}
H_c^{'}(T)=H_c^{'}(0)\bigg[1-\frac{e^{2\gamma}}{3}\;\tau'^{2}
\bigg]\doteq H_c^{'}(0)\big[1-1.06\;\tau'^{2} \big] \;,\;T \ll
T^{'}_c \;.
\end{eqnarray}
From the last formulae one sees qualitative differences between
the critical magnetic fields $H_c(T)$ and $H_c^{'}(T)$
corresponding to the unconventional solution and to the standard
solution resp. of the BCS theory. The curve $H_c(T)$ is convex for
$\tau \in \big(0,\frac{1}{2}\big)$ and concave for $\tau \in
\big(\frac{1}{2},1\big)$ while $H_c^{'}(T)$ of the standard
solution is a convex curve for the full temperature interval
$\tau' \in (0,1)$.

\begin{figure}[h]
\bigskip{}
\centering\resizebox{0.6\textwidth}{!}{\includegraphics{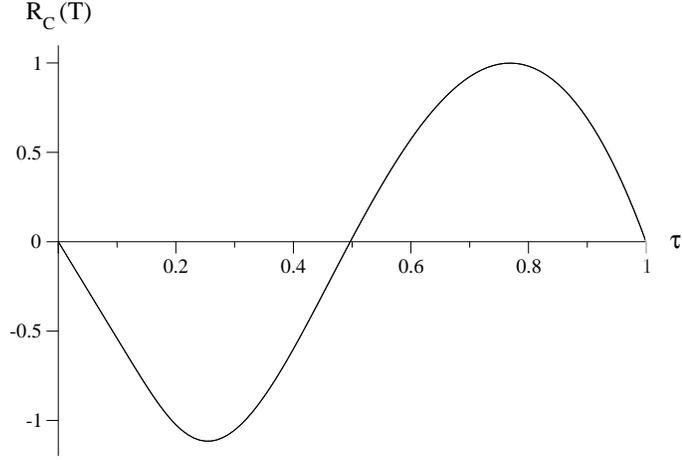}
\par} \caption{The specific heat anomaly corresponding to the unconventional
solution of the BCS theory.} \label{fig5}
\bigskip{}
\end{figure}

We now calculate the specific heat anomaly from the difference
(\ref{6.6}) to get the result
\begin{eqnarray}\label{6.16}
\frac{C_s(T)-C_n(T)}{C_n(T)}=\frac{3}{8}g N(0) \left \{
\frac{2}{\tau^2}\frac{\eta^4(1-\eta^2)}{\eta^2+\tau-1}-\frac{1}{2}\tau
\varphi(\eta) \right \} \;,
\end{eqnarray}
where
\begin{eqnarray}\label{6.17}
C_n(T)=V\frac{2\pi^2}{3}N(0)k_B^2T
\end{eqnarray}
is the specific heat of an ideal electron gas. The temperature
behavior of the specific heat anomaly (\ref{6.16}) plotted as the
ratio
\begin{eqnarray}\label{6.18}
R_C(\tau)=\frac{8}{3gN(0)}\frac{C_s(T)-C_n(T)}{C_n(T)}
\end{eqnarray}
is shown on the Fig. \ref{fig5}.
\\
The specific heat anomaly (\ref{6.16}) has the following behavior
in the limits
\begin{eqnarray}\label{6.19}
\frac{C_s(T)-C_n(T)}{C_n(T)}=\frac{27}{8}g N(0) (1-\tau)
\;,\;T_c-T \ll T_c
\end{eqnarray}
and
\begin{eqnarray}\label{6.20}
\frac{C_s(T)-C_n(T)}{C_n(T)}=-\frac{27\zeta(3)}{16}g N(0)\tau
\;,\;T \ll T_c \;.
\end{eqnarray}
This behavior of the specific heat anomaly is qualitatively
completely different from that corresponding to the standard
solution \cite{fetter} of the BCS theory
\begin{eqnarray}\label{6.21}
\frac{C_s^{'}(T)-C_n(T)}{C_n(T)}=\frac{12}{7\zeta(3)}+
O\big[(1-\tau)^2\big]\doteq 1.43+O\big[(1-\tau)^2\big] \;,\;T_c-T
\ll T_c
\end{eqnarray}
and
\begin{eqnarray}\label{6.22}
\frac{C_s^{'}(T)-C_n(T)}{C_n(T)}=-1+O\big(e^{-\frac{1}{\tau}}\big)\;,\;T
\ll T_c \;.
\end{eqnarray}
The specific heat $C_s(T)$ corresponding to the unconventional
superconductivity (\ref{6.16}) is a continuous function of $T$ at
$T=T_c$, as is seen from (\ref{6.19}), however, with a
discontinuous derivative at $T=T_c$ in a sharp contradistinction
to the standard case. Thus the unconventional solution considered
here gives rise to a superconducting phase transition of the third
order. From the results (\ref{6.8})-(\ref{6.20}) it is evident
that the properties of superconducting states associated with two
different inequivalent representations of the canonical
anticommutator ring of the electron field operators (\ref{2.15})
are qualitatively completely different.

Despite of the fact that we did not have any ambitions to explain some
experimental data, but only to study the theoretical consequences
coming from different inequivalent representations of the
canonical anticommutator ring (\ref{2.15}) within the framework of
the BCS theory, we nevertheless mention a resemblance of the
theoretical results (\ref{6.16})-(\ref{6.20}) to existing
experimental data. The specific heat anomaly on the Fig.
\ref{fig5} for $\tau \in \big(\frac{1}{2},1\big)$ is similar to
the specific heat anomalies of cuprate HTS \cite{junod1}. The
shape of the specific heat anomaly as $T$ approaches to zero, as
given by (\ref{6.20}), tells us that the specific heat $C_s(T)$ in
this novel superconducting state behaves as a polynomial of $T$,
i.e., similarly as electronic components of ordinary metals in the
normal state. This behavior is again, perhaps accidentally,
consistent with experimental data of cuprates HTS
\cite{junod1,junod2} and is completely different from the behavior
of LTS described by the standard solution of the BCS theory. We
also recall that the specific heat jump at $T_c$ for the standard
solution of the BCS theory (\ref{6.21}) is given by a universal
number which is fairly consistent with experimental data for LTS.
However, the novel superconducting state discussed here relates
the specific heat anomaly (\ref{6.16}) to the normal specific heat
$C_n(T)$ by a material parameter $gN(0)$. This feature seems to be
again in a correspondence with experimental data of HTS
\cite{junod2}. For example, the specific heat anomaly of
$\mathrm{YBa_2Cu_3O_x}$ which ranks highest among HTS, does not
exceed $5\%$ of the normal specific heat. The situation for
$\mathrm{Bi_2Sr_2CaCu_2O_x}$ is about three times worse. Thus it
seems that the properties of the unconventional solution to the
BCS Hamiltonian investigated in this paper may have some relevance
for understanding of the superconductivity mechanism in HTS.

The system of electrons under consideration, however, stabilizes
in a phase corresponding to the minimal Helmholtz free energy.
Thus, for completeness, we analyze the stability of the novel
superconducting phase resulting from the unperturbed Hamiltonian
$H_{03}$ with respect to the standard BCS phase arising from the
unperturbed Hamiltonian $H_{02}$. The Helmholtz free energy in
each of the superconducting phases is determined by the critical
magnetic fields $H_c(T)$ and $H^{'}_c(T)$ as given by the relation
(\ref{6.8}). The critical magnetic field $H_c(T)$ for the novel
superconducting phase is given by the relations
(\ref{6.9})-(\ref{6.10}). The temperature behavior of the critical
magnetic field $H^{'}_c(T)$ corresponding to the standard
superconducting phase can be approximated with a high accuracy by
the formula
\begin{eqnarray}\label{6.23}
H^{'}_c(T)=H^{'}_c(0)(1-\tau'^2)\;,
\end{eqnarray}
where $\tau'=\frac{T}{T^{'}_c}$ and $H^{'}_c(0)$ is given by
(\ref{6.13}). It is convenient to introduce the "reduced"
Helmholtz free energy difference $\Delta f$ defined by
\begin{eqnarray}\label{6.24}
\Delta f \equiv \frac{8\pi(F_S-F_N)}{VH^{'2}_c(0)}
\end{eqnarray}
which has the following expression
\begin{eqnarray}\label{6.25}
\Delta f=-\frac{H^2_c(0)}{H^{'2}_c(0)}
\bigg[\eta^4-\frac{1}{2}\tau^3\varphi(\eta)\bigg]
\end{eqnarray}
for the novel superconducting phase and
\begin{eqnarray}\label{6.26}
\Delta f^{'}= -(1-\tau'^2)^2
\end{eqnarray}
for the standard BCS result. The novel superconducting phase is
preferable over the standard phase provided that the following
inequality is satisfied:
\begin{eqnarray}\label{6.27}
\Delta f < \Delta f^{'}\;.
\end{eqnarray}
The last inequality is always satisfied if
$\frac{H_c(0)}{H^{'}_c(0)}>0$, i.e., if the material parameters
$\omega_D, \epsilon_F, g$ and the thickness of the layers $d$
satisfy the following condition
\begin{eqnarray}\label{6.28}
T_c>e^{-\gamma}\sqrt{\frac{6}{gN(0)}}\;T^{'}_c\;,
\end{eqnarray}
where $T_c$ and $T^{'}_c$ should be expressed in terms of material
parameters as given by (\ref{5.12})-(\ref{5.14}) and (\ref{4.23})
respectively.

\begin{figure}[h]
\bigskip{}
\centering\resizebox{0.6\textwidth}{!}{\includegraphics{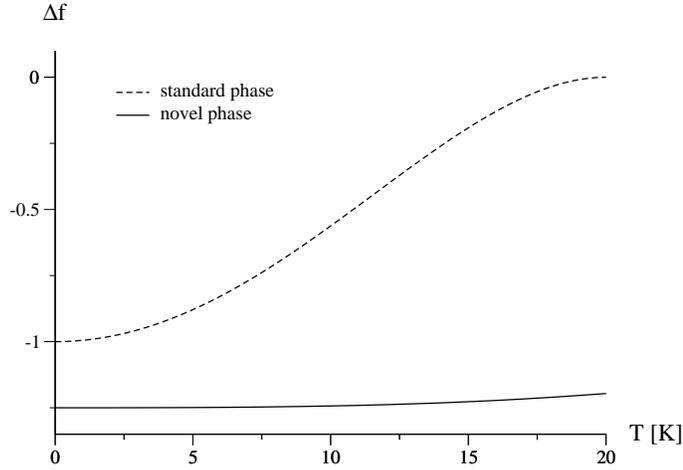}
\par} \caption{The comparison of Helmholtz free energies corresponding
to the standard and the novel solution to the BCS theory for
$T_c=100$ K, $T^{'}_c=20$ K and $gN(0)=0.1$.} \label{fig6}
\bigskip{}
\end{figure}

Since we have the same Hamiltonian responsible for both the novel
superconducting phase and for the standard BCS solution in two
inequivalent representations of the anticommutator ring of
electron field operators we may assume that the values of the
coupling constant $g$ and the density of electron states $N(0)$ at
the Fermi surface are the same as those experimentally found for
LTS \cite{parks}. Therefore, the values $gN(0)\in(0.1,0.5)$
\cite{parks}. For the reasonably chosen values \mbox{$T_c=100$ K},
$T^{'}_c=20$ K and $gN(0)=0.1$ the numerical results of
temperature dependence of the reduced Helmholtz free energies
$\Delta f$ are shown on the Fig. \ref{fig6}.
\\
Fig. \ref{fig6} clearly shows that the novel superconducting phase
associated with the third inequivalent representation of the
anticommutator ring of electron field operators of the same BCS
Hamiltonian is energetically preferable with respect to its
standard superconducting phase.

\newpage
\section{Summary}\label{sec7}
In applications of quantum field theory of many particle systems
one has intuitively believed that a given Hamiltonian $H$
determines uniquely a single Gibbs state of a system at given
values of thermodynamic variables like temperature $T$, volume $V$
and the number $N$ of particles. Such a belief would be completely
correct provided that the canonical commutation or anticommutation
relations of field operators had a unique representation. In the
thermodynamic limit $V \to \infty$, $N \to \infty$, $\frac{N}{V}
\to \mathrm{constant}$, one deals, in fact, with systems having
the infinite number of degrees of freedom. In this case, as it has
been pointed out by Haag \cite{haag}, the canonical commutation or
anticommutation relations of field operators have no longer unique
solutions, i.e., they have several different inequivalent
representations. This fact has to be taken into account in all
discussions concerning the quantum field theory of many particle systems.
Namely, with each inequivalent representation of commutation or
anticommutation relations one has to associate the corresponding
"matrix form" of the Hamiltonian $H$ leading to the corresponding
grand canonical partition function $Z$ and the grand canonical
potential $\Omega$ which fully specify the Gibbs state of the
system. This argument indicates the fact that the system can have
as many Gibbs states as many inequivalent representations of the
canonical commutator or anticommutator ring exist for the given
Hamiltonian $H$.

To our best knowledge there is no systematic method for a
classification of all inequivalent representations of the
canonical commutator or anticommutator ring of field operators
entering a given Hamiltonian $H$, e.g., as the method one has for
the classification of the irreducible representations of Lie
algebras. The only known way how one selects an inequivalent
representation of the canonical commutator or anticommutator ring
of field operators entering a given Hamiltonian $H$ is to 
select a suitable unperturbed Hamiltonian $H_0$ which can be
exactly diagonalized.

In the present work we have explicitly constructed three
inequivalent representations of the canonical anticommutator ring
of electron field operators entering the Hamiltonian of the BCS
theory of superconductivity. Each inequivalent representation has
been constructed by selecting the unperturbed
Hamiltonian $H_{01}$, $H_{02}$ or $H_{03}$. Each inequivalent
representation specifies the corresponding Gibbs state of the
system. The physical system, however, stabilizes in a Gibbs state
having the minimal Helmholtz free energy at given values of the
thermodynamic variables like temperature $T$, volume $V$ and the
particle number $N$. From this reason it is important to
search for all relevant inequivalent representations of the
canonical anticommutator ring of field operators entering the
Hamiltonian of the system. The normal state and the standard
superconducting state of the BCS theory corresponding to two
inequivalent representations have been generally known long time
ago. We have constructed the third inequivalent representation
which has not been known till now. This representation can,
perhaps, be relevant for describing the superconducting properties
of HTS.

\section{Acknowledgements}\label{sec6}
The authors are very grateful for many elucidating discussions
with Professors C. Cronstr\"om, I. Huba\v{c} and R. Hlubina.

\end{document}